\documentclass[useAMS,usenatbib,letterpaper]{mn2e}
\usepackage[totalwidth=480pt,totalheight=680pt]{geometry}
\usepackage{graphicx,amssymb,amsbsy}
\input psfig

\newcommand{\be}{\begin{equation}}
\newcommand{\ee}{\end{equation}}
\def\lta{\,\raise 0.3 ex\hbox{$ < $}\kern -0.75 em
 \lower 0.7 ex\hbox{$\sim$}\,}
\def\gta{\,\raise 0.3 ex\hbox{$ > $}\kern -0.75 em
 \lower 0.7 ex\hbox{$\sim$}\,} 
\newcommand{\zhat}{{\hat z}} 
\newcommand{\norm}{ {\cal N}} 
\newcommand{\kepler}{{\it Kepler}\hskip0.5em}
 
\newcommand{\ratqs}{{\cal R}_{\rm qs}}  
\newcommand{\ndi}{\chi} 
\newcommand{\mumass}{{\mu}} 
\newcommand{\omegaorb}{ \Omega_{\rm orb}} 
\newcommand{\qpot}{ Q_\Phi} 

\title[On the Stability of Extrasolar Planetary Systems] 
{On the Stability of Extrasolar Planetary Systems\\
and other Closely Orbiting Pairs} 
\author[Adams \& Bloch]{Fred C. Adams$^{1,2}$ and Anthony M. Bloch$^{3}$\\
$\,$\\ 
$^1$Physics Department, University of Michigan, Ann Arbor, MI 48109\\
$^2$Astronomy Department, University of Michigan, Ann Arbor, MI 48109\\
$^3$Mathematics Department, University of Michigan, Ann Arbor, MI 48109 }

\begin{document} 

\date{October 2014} 

\pagerange{\pageref{firstpage}--\pageref{lastpage}} \pubyear{2014}

\maketitle

\label{firstpage}

\begin{abstract}
This paper considers the stability of tidal equilibria for planetary
systems in which stellar rotation provides a significant contribution
to the angular momentum budget. We begin by applying classic stability
considerations for two bodies to planetary systems --- where one mass
is much smaller than the other. The application of these stability
criteria to a subset of the \kepler sample indicates that the majority
of the systems are not in a stable equilibrium state.  Motivated by
this finding, we generalize the stability calculation to include the
quadrupole moment for the host star.  In general, a stable equilibrium 
requires that the total system angular momentum exceeds a minimum
value (denoted here as $L_X$) and that the orbital angular momentum of
the planet exceeds a minimum fraction of the total. Most, but not all,
of the observed planetary systems in the sample have enough total
angular momentum to allow an equilibrium state. Even with the
generalizations of this paper, however, most systems have too little
orbital angular momentum (relative to the total) and are not in an
equilibrium configuration.  Finally, we consider the time evolution of
these planetary systems; the results constrain the tidal quality
factor of the stars and suggest that $10^6\lta{Q_\ast}\lta10^7$.
\end{abstract}

\begin{keywords}
binaries: close --- planets and satellites: dynamical evolution and
stability --- planetary systems --- stars: kinematics and dynamics
\end{keywords} 

\section{Introduction} 
\label{sec:intro} 

For two-body systems that include both rotational and orbital motion,
the conditions required for the existence of a stable tidal
equilibrium state have been determined (\citealt{darwin1,darwin2};
\citealt{counselman,hut1980}).  This previous work shows that if the
system can dissipate energy, for example through the action of tides,
it can evolve in three possible ways: [1] The orbit of the secondary
can move outward toward an unbound state, albeit at an ever-decreasing
rate. [2] The orbit can decay inward and eventually collide with the
primary. [3] The orbit can approach an equilibrium configuration
characterized by equal periods for the orbit and spins of both bodies,
circularization of the orbits, as well as alignment of the three
angular momentum vectors.

In recent years, this classic problem has been the subject of renewed
interest because it plays a role in a number of astrophysical
contexts: Hot Jupiters can be destroyed via tidal dissipation by
subgiants \citep{schlaufman}, and can spin up their parental stars as
they spiral inward \citep{zhang2014}. The tidal destruction of
extrasolar planets -- or lack thereof -- can be used to place
constraints on the tidal quality factor of the host stars
\citep{penev2012}. Similarly, a mass limit can be derived for
hypothetical moons orbiting Jovian exoplanets \citep{barnes2002}. Many
extrasolar planetary systems with Hot Jupiters (apparently) do not
have an equilibrium state, and this complication changes the required
description of their subsequent tidal evolution \citep{levrard}. The
alignment and evolution of planetary obliquity can also affect the
habitability of planets \citep{heller2011}. In addition to exoplanets,
this issue arises in many other astronomical systems, including
non-spherical binary asteroids \citep{scheeres02,bellerose,scheeres09}, 
common envelope evolution of binary stars \citep{taam2000}, the
evolution of compact binary systems \citep{postnov}, and period gaps
in binary millisecond pulsars \citep{taamking}.

This paper has two coupled goals. The first goal is to apply existing
stability criteria to the planetary candidates discovered by the
\kepler mission. The second goal is to generalize the stability
criteria. Toward these ends, we first review the basic approach used
in previous work \citep{counselman,hut1980}, and introduce notation
appropriate for the extrasolar planetary systems of interest (Section
\ref{sec:basic}). We apply these stability criteria to to a subset of
the \kepler sample for which stellar rotational periods are measured,
and find that the majority of systems are not in a tidal equilibrium
state (Section \ref{sec:apply}).  Motivated by this finding, as well
as the applications outlined above, Section \ref{sec:quad} generalizes
the classic problem of the stability of two-body systems containing
spin angular momentum by augmenting the stellar potential to include a
quadrupole term. However, this correction is small and the majority of
the observed systems are not in a tidal equilibrium state.  We deduce
that the systems must still be evolving dynamically and consider the
corresponding implications in Section \ref{sec:evolve}; these results
imply constraints on the tidal quality factor $Q_\ast$ of the star.
The paper concludes, in Section \ref{sec:conclude}, with a summary and
discussion of our results.

\section{Stability of Planetary Systems Including Stellar Spin} 
\label{sec:basic} 

This section considers the equilibrium state of a two-body system
consisting of a star and a single planet. To find this state, we need
to extremize the system energy $E$ subject to the constraint that the
total angular momentum is constant. This treatment is parallel to that
of previous work on binary stars \citep{hut1980,counselman}, and 
analogous to more general treatments of energy methods in stability 
problems \citep{wang1991,simo}. 

In this system, both the energy and angular momentum budgets have
contributions from three sources: the orbit, the spin of the star, and
the spin of the planet. For the systems of interest here, the planets
are small and have relatively little spin angular momentum; we thus
reduce the problem by working in the limit where the planetary angular
momentum vanishes. The orbit of the star-planet system can be
described by the standard six orbital elements. In this case, however,
we are only interested in the three variables $(a,e,i)$ because the
remaining ones can be averaged over; in other words, they only play a
role on short timescales.

The star has moment of inertia $I$ and spin angular momentum vector
\be
{\bf S} = {S}\zhat \equiv I\Omega \zhat\,,
\ee 
where the $\zhat$ direction is coincident with the pole of the star
and $\Omega$ is the angular speed of the star.  Following the
treatment of \cite{hut1980}, the total angular momentum of the system
is conserved and is given by 
\be
{\bf L} (a,e,i,\Omega) = {\bf h} + I \Omega \zhat \,,
\ee
where ${\bf h}$ is the orbital angular momentum, with magnitude $h$ 
given by 
\be
h^2 = \mumass^2 G (M+m) a (1-e^2) \,,
\ee
where $M$ is the stellar mass, $m$ is the planetary mass, and
$\mumass$ is the reduced mass\footnote{Note that many different  
notations exist for the reduced mass: Here we follow \cite{goldstein}
and use $\mu$; compare with \cite{md} who use $\mu^\ast$ 
and \cite{morby} who uses $\mu_1$.} defined by  
\be
\mumass = {m M \over M + m} \,. 
\ee
The direction of the orbital angular momentum vector is 
defined so that 
\be
{\bf h} \cdot \zhat = h \cos i \,. 
\ee
The energy of the system is the sum of the orbital and spin energies, 
and is given by  
\be
E = - {GMm \over 2a} + {1 \over 2} I \Omega^2 \,. 
\label{energy} 
\ee  
Without loss of generality, we can define the direction of 
the orbital angular momentum vector so that 
\be
{\bf h} = (h \sin i, 0, h \cos i) \,,
\label{angmomvec} 
\ee
which thus defines the ${\hat x}$-axis. The total angular  
momentum vector ${\bf L}$ can then be written 
\be
{\bf L} = (h \sin i, 0, h \cos i + I \Omega) \,.
\label{angmoment} 
\ee

\subsection{Extremum of the Energy} 

The basic problem is to find the extremum of the energy $E$ given by
equation (\ref{energy}) subject to the constraint that the total
angular momentum ${\bf L}$ (given by equation [\ref{angmoment}]) is
constant. The energy is a function of four variables, including the
semimajor axis $a$, the eccentricity $e$, the inclination angle $i$,
and the spin rate $\Omega$ of the star. The mass $M$ of the star, mass
$m$ of the planet, and moment of inertia $I$ are considered fixed.
Since the angular momentum has only two nonzero components, we need
two Lagrange multipliers (equivalently, the Lagrange multiplier is a
two-dimensional vector). We thus introduce the two unknown quantities
$(\lambda_x,\lambda_z)$. For each variable $x_k$, we get an
optimization condition of the form 
\be
{\partial E \over \partial x_k} + 
\lambda_x {\partial \over \partial x_k} (h \sin i) + 
\lambda_z {\partial \over \partial x_k} (h \cos i + I \Omega) = 0 \,,
\ee
where the $x_k$ are the four variables $(a,e,i,\Omega)$.

The above approach yields four equations that specify the 
tidal equilibrium state: 

[1] For the semimajor axis $a$, the condition becomes 
\be
{G M m \over a} + \left[ \lambda_x \sin i 
+ \lambda_z \cos i \right] h = 0 \, . 
\label{aconstraint} 
\ee

[2] For the eccentricity $e$, optimization takes the form 
\be
\left[ \lambda_x \sin i + \lambda_z \cos i \right] 
{e \over 1 - e^2} = 0 \,.
\label{econstraint} 
\ee

[3] For the inclination angle $i$, we obtain 
\be
\lambda_x h \cos i - \lambda_z h \sin i = 0 \,. 
\label{iconstraint} 
\ee

[4] And finally for the rotation rate $\Omega$ of the star, 
the constraint can be written 
\be
I \Omega + \lambda_z I = 0 \,.
\label{oconstraint} 
\ee

The eccentricity equation (\ref{econstraint}) implies that $e=0$, 
and the other equations have solution $i=0$, $\lambda_x=0$, and 
$\lambda_z=-\Omega$. The remaining condition thus becomes
\be
{GMm \over a} = \Omega h = \Omega {Mm\over M+m} 
\left[G (M+m) a \right]^{1/2}  \,.
\ee
As a result, the spin rate of the star must match 
the orbital angular velocity of the planet, 
\be
\Omega = \left[{G (M+m) \over a^3} \right]^{1/2}\,.
\label{sync} 
\ee
The total angular momentem is then given by 
\be
L = h + I \Omega = 
{Mm G^{2/3} \over (M+m)^{1/3}} \Omega^{-1/3} + I \Omega \,.
\label{angcurve} 
\ee
Given this expression, we see that $L\to\infty$ in both limits 
$\Omega \to 0$ and $\Omega \to \infty$. As a result, there exists a
critical value of the total angular momentum $L_X$, such that no
equilibrium exists for smaller values. This critical angular momentum
is determined by finding the minimum of equation (\ref{angcurve}) as 
a function of $\Omega$ and is given by 
\be 
L_X = {4 \over 3} \left[ {3 I (Mm)^3 G^2 \over M+m} \right]^{1/4}\,.
\label{deflx} 
\ee
At the critical point, the orbital angular momentum makes up 
three fourths of the total, whereas the stellar spin represents 
the remaining one fourth. This result is in agreement with that 
obtained earlier \citep{hut1980}. 


\subsection{Second Variation}

In order for the system to be in equilibrium, the extremum found in
the previous subsection must be a minimum of energy (rather than a
maximum). Strictly speaking, the maximum only destabilizes in the
presence of dissipation \citep{bloch1994}; in the present application
we expect dissipation over the long term, although is can be rather
weak (see Section \ref{sec:evolve}). If we use conservation of angular
momentum, 
\be
{\bf L} = {\bf h} + I \Omega \zhat\,,
\ee
we can write the energy in the form 
\be
E = - {GMm \over 2a} + {1 \over 2I} 
\left[L^2 + h^2 - 2 L h \cos \theta \right] \,,
\label{secondenergy} 
\ee
where $\theta$ is the angle between the total angular momentum 
${\bf L}$ and the orbital angular momentum ${\bf h}$. Note that
$\theta$ is not the same as the inclination angle defined earlier, 
but we can use $\theta$ as the third variable and find that 
$\theta=i=0$ in the equilibrium state \citep{hut1980}. 

After some algebra, the second derivatives, evaluated at the 
equilibrium conditions, have the forms 
\be
{\partial^2 E \over \partial a^2} = {GMm \over 4a^3} 
\left[ -3 + {\mumass a^2 \over I} \right] \,, 
\label{ddedaa} 
\ee
\be 
{\partial^2 E \over \partial e^2} = {GMm \over a} \,, 
\label{ddedee} 
\ee
and 
\be 
{\partial^2 E \over \partial \theta^2} = {GMm \over a} 
\left[ 1 + {\mumass a^2 \over I} \right] \,. 
\label{ddedhh} 
\ee
Since all of the off-diagonal terms vanish at the equilbrium state,
these three second-partial-derivatives also define the eigenvalues of
the relevant Hessian matrix \citep{hesse}. The second two expressions
are manifestly positive. The only nontrivial constraint required for
stability is that the first expression (from equation [\ref{ddedaa}])
is positive, which implies 
\be
{\mumass a^2 \over I} > 3 \,.
\ee 
At the critical point, the orbital angular momentum has the form 
$h = \mumass a^2 \Omega$, so the above constraint can be written in 
the alternate form  
\be
h = \mumass a^2 \Omega > 3 I \Omega \,. 
\label{hsthree} 
\ee
In other words, the orbital angular momentum must be three times
larger than the spin angular momentum in order for the system to 
be in its stable equilibrium state (in agreement with the results 
of \citealt{hut1980} in the limit where the companion has no spin).
Notice also that since $L = h + I\Omega$, the above condition can 
be written in the alternate form $h > 3L/4$. 

\subsection{Stability and Instability} 

\begin{figure} 
\centerline{\psfig{figure=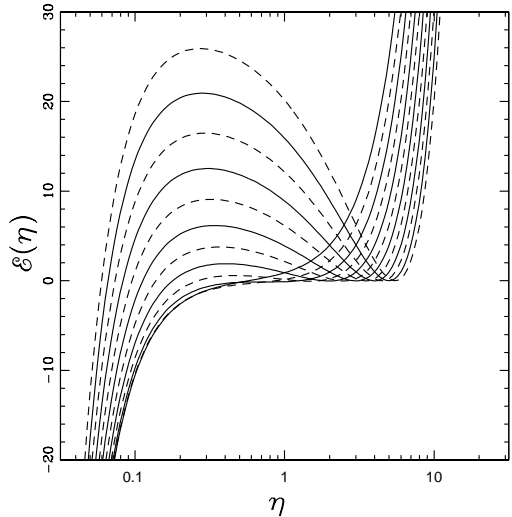,width=8.5cm}}
\caption{Total system energy as a function of dimensionless orbital
angular momentum. Energy curves are shown for a range of total angular 
momenta $\ell=L/L_X$, which are equally spaced. From bottom to top, the 
solid curves correspond to $\ell$ = 1, 2, 3, 4, and 5; the dashed curves 
show $\ell$ = 3/2, 5/2, 7/2, 9/2, and 11/2. Planetary systems (with no 
external torques) conserve angular momentum and must follow these paths. 
In systems that start to the left of the local maximum, the planet spirals 
inwards; in systems that start to the right of the maximum, the planet 
spirals outward until it reaches the local minimum of energy. } 
\label{fig:curves} 
\end{figure}

The meaning of the tidal equilibrium state dervied above can be
illustrated by considering the system energy as a function of angular
momentum, or, equivalently, semimajor axis.  First we define the
dimensionless energy and orbital angular momentum according to
\be
{\cal E} \equiv {E \over (L_X^2/2I)} \qquad {\rm and} \qquad 
\eta \equiv {h \over L_X} \,. 
\ee
Next we specialize to the case where the spin of the star is aligned
with the direction of the orbit and the eccentricity vanishes (note
that $i$ = 0 = $e$ is necessary for equilibrium).  The energy from
equation (\ref{secondenergy}) then has the form
\be
{\cal E} = - {27 \over 256} {1 \over \eta^2} + 
\left( \ell - \eta \right)^2 \,,
\label{reduced} 
\ee
where we have defined $\ell \equiv L/L_X$. We then plot the energy as
a function of (dimensionless) orbital angular momentum, as shown in
Figure \ref{fig:curves}. For $\ell \le 1$, no equilbria are possible,
and the energy is a monotonic function of $\eta$. For $\ell > 1$, the
energy curve has a stable equilibrium point at some value $\eta_{+}>1$
and an unstable equilibrium point at $\eta_{-}<1$. Planetary systems
with a given value of total angular momentum ($\ell$) will fall on the 
corresponding energy curve in this diagram. If their location falls to 
the left of the local maximum, the planet can spiral inward and would 
eventually be accreted. 

\begin{figure} 
\centerline{\psfig{figure=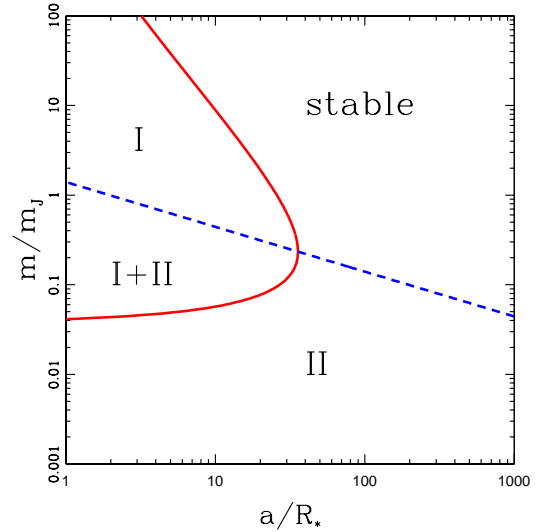,width=8.5cm}}
\caption{Regions of stability and instability for systems where the 
primary has solar properties. The horizontal axis marks the semimajor
axis of the orbit (in units of solar radii) and the vertical axis
marks the mass of the secondary (in Jupiter masses). In the region to
the left of the solid red curve (region I), the systems fail to have
enough total angular momentum for tidal equilibrium ($L<L_X$); in the
region below the blue dashed line (region II), systems fail to have
enough orbital angular momentum ($h<3S$). Stable orbits fall in the
upper right portion of the plane. } 
\label{fig:maplane} 
\end{figure} 

Next we want to delineate the parts of parameter space that lead to
stable and unstable configurations. The binary systems considered here
have a large number of parameters, including the masses $m$ and $M$,
the stellar rotation rate $\Omega$, the moment of inertia $I$, and the
orbital elements $(a,e,i)$ of the secondary.  Since we are primarily
interested in planetary systems, we can fix the stellar properties,
which are chosen (for now) to be those of the Sun. For the sake of
definitness, we also consider circular orbits in the plane $(e=0=i)$.
Some systems will have smaller stars and different rotation rates, but
these differences have less dynamic range than those of the planet
masses and semimajor axes. The parameter space thus reduces to $(a,m)$. 

With the above specifications, we plot the regions of stability in the
$m$-$a$ plane in Figure \ref{fig:maplane}.  The semimajor axes $a$ are
given in units of the stellar radius and the planet masses are
expressed relative to the mass of Jupiter. This diagram shows that
planetary systems can be unstable for different reasons. The area to
the left of the solid red curve delineates the parameter space for
which the systems have too little total angular momentum $(L<L_X)$ so
that no equilibrium state is possible; this condition is denoted here
as type-I instability. The area below the dashed blue line delineates
the parameters for which the systems have too little orbital angular
momentum relative to the spin (so that $h<3S$); this condition is
denoted as type-II instability. Note that systems can fail both
requirements and be unstable for two reasons (for parameters in the
middle left part of the plot). The upper right portion of the diagram
delineates the parameters for which the systems are stable. Note that
systems with Jovian planets are susceptible to type-I instability (not
enough total angular momentum), whereas smaller planets are more
likely to suffer type-II instability (not enough orbital angular
momentum). In this context, small planets are those with lower masses
than Neptune. Note that for smaller stars (and/or slower rotation rates), 
the region of type-I instability will be larger. 

\section{Application to Kepler Planets} 
\label{sec:apply} 

This section uses the stability criteria outlined above to analyze a
subset of the \kepler sample of extrasolar planet candidates
\citep{batalha}. In order to apply the stability conditions, the
rotation rates of the host stars must be known. Toward that end, we
use the results of \cite{mcquillan}, who detected rotational periods
for 797 of the stars that host \kepler objects of interest. This set
of systems is reduced further by eliminating known eclipsing binaries,
previously published blended objects, systems that are likely to be
eclipsing binaries, and systems whose centroid motions indicate
rotation and transit modulation on different stars (for further
detail, see \citealt{mcquillan}). With this reduction, the sample of
contains 738 planetary systems. Within the sample, we consider only
the innermost planetary candidates, which are then assumed to be real
with their reported radii and orbital elements. For the sake of
definiteness, we convert planetary radii to masses using the relation
$m = M_{\earth} (R_p/R_{\earth})^{2.1}$ \citep{lissauer}, which is
appropriate for smaller planets ($m < 150 M_{\earth}$; for example, see
\citealt{weiss} and references therein). The orbital eccentricities
are not generally measured and are set to zero for this analysis. For
the stars, in addition to their reported properties, we assume that
the dimensionless moment of inertia has a single value $\ndi=I/(MR^2)$
= 0.10, which is intermediate between that of a fully convective
$n=3/2$ polytrope and a fully radiative $n=3$ polytrope (e.g., see
Figure 2 of \citealt{batygin}). Note that the correction for nonzero
eccentricity is ${\cal O}(e^2)$ and that the moment of inertia
dependence has the form $L_X \propto I^{1/4}$ (see equation
[\ref{deflx}]), so that the results presented below are relatively
insensitive to these approximations.
 
\begin{figure} 
\centerline{\psfig{figure=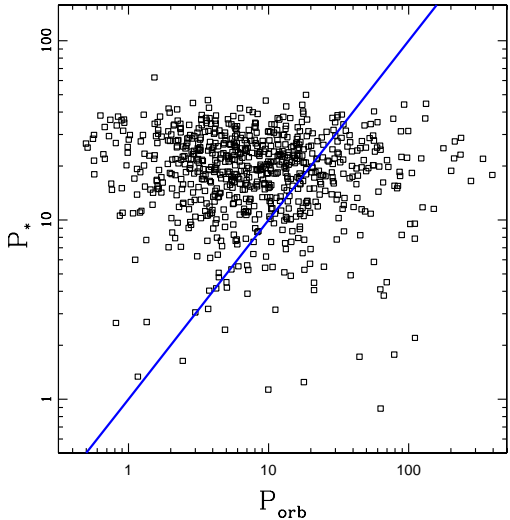,width=8.5cm}}
\caption{Stellar rotation period versus orbital period 
of the innermost planet for \kepler candidates (see
\citealt{batalha,mcquillan}). The blue line depicts equal 
periods. Note that the systems do not generally exhibit 
synchronicity.} 
\label{fig:pstarorb} 
\end{figure}

First we plot the stellar rotation period versus the orbital period of
the innermost planet, as shown in Figure \ref{fig:pstarorb} (which is
analogous to Figure 2 of \citealt{mcquillan}). This figure shows
immediately that the observed planetary systems are not synchronous in
general, and hence are not in stable equilibrium states. The diagonal
blue line in the figure shows the locus of equal periods. Note that
the data fall on both sides of this line of synchronicity. Although
more points fall above the line, the transits are more likely to be
observed for shorter orbital periods, so selection effects could
account for this asymmetry. Another interesting feature of Figure
\ref{fig:pstarorb} is that it shows no apparent correlation between
the stellar rotation period and the orbital period. In other words,
the data points do not cluster around the expected line of
synchronicity, but rather appear to be completely independent.

\begin{figure} 
\centerline{\psfig{figure=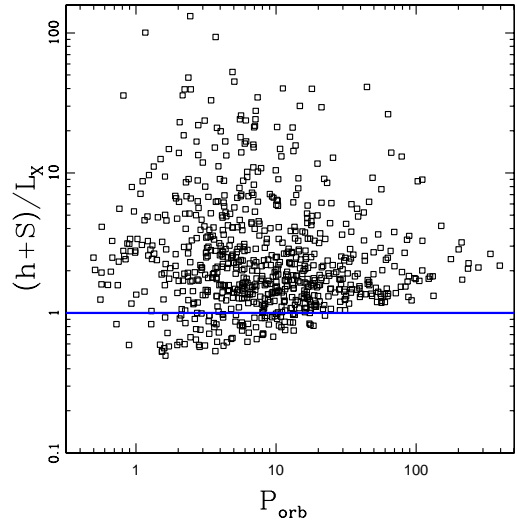,width=8.5cm}}
\caption{Ratio of the total angular momentum to the critical value,
plotted as a function of orbital period. Only systems with sufficient 
angular momentum, those above the blue line in the figure, have a 
tidal equilibrium state. The area below the line corresponds to 
region I instability from Figure \ref{fig:maplane}. The 85 systems 
(out of 738) that fall below the horizontal blue line have no tidal 
equilibrium state. } 
\label{fig:critang} 
\end{figure}

Given that the planetary systems are not in a co-rotating state, which
is required for equilibrium, the next step is to determine if such an
equilibrium exists. The result is shown in Figure \ref{fig:critang}.
The vertical axis plots the ratio of the total (spin plus orbit)
angular momentum of the system to the minimum value $L_X$ needed for
the existence of an equilibrium state (see equation [\ref{deflx}]).
This ratio is plotted versus orbital period of the planet. Even though
most systems are not in tidal equilibrium (as indicated by Figure
\ref{fig:pstarorb}), an equilibrium state does exist for the majority
of the cases (most systems lie above the critical line).  Nonetheless,
85 systems (out of 738 total) fall below the critical blue line in
Figure \ref{fig:critang}. These systems have no accessible equilibrium
state and are subject to type-I instability (see Figure
\ref{fig:maplane}); these planets must eventually either spiral inward
or outward.

\begin{figure} 
\centerline{\psfig{figure=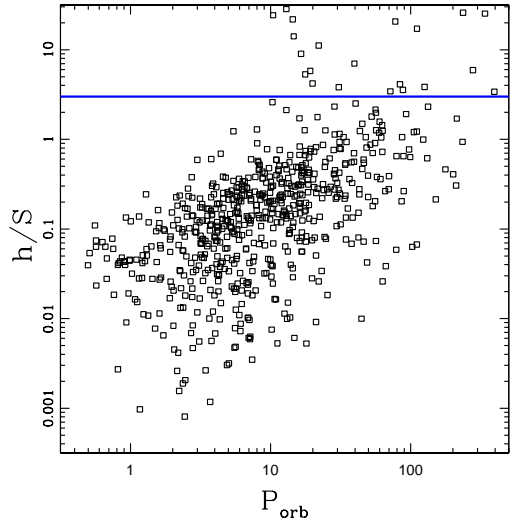,width=8.5cm}}
\caption{Ratio of the orbital angular momentum $h$ to the spin  
angular momentum $S$ of the star, plotted here plotted as a function
of orbital period. Only systems with $h/S > 3$, those above the blue
line in the figure, can reside in their tidal equilibrium state. The 
area below the line corresponds to region II instability in Figure 
\ref{fig:maplane}. This plot shows that the majority of planetary
systems in the sample are not in equilibrium. Only the systems for
which the equilbrium state exists (those with $L>L_X$) are included in
the plot. }
\label{fig:hsratio} 
\end{figure}

Although a sizable majority (653 out of 738) of the planetary systems
in the sample have enough angular momentum for a tidal equilibrium
state (Figure \ref{fig:critang}), only a small fraction of the systems
are in a synchronous state within the observational uncertainties
(synchronicity is one of the conditions to be in stable equilibrium).
Another requirement to be in stable equilibrium is for the orbital
angular momentum $h$ to represent a sufficiently large fraction of the
total $L$. For the case of no quadrupole ($q\to0$), this condition can
be written as $h>3L/4$, or, equivalently, $h>3S$.  As expected, the
ratio $h/S$ is generally smaller than indicated by this requirement,
as shown in Figure \ref{fig:hsratio}. In fact, most of the systems
have far too little orbital angular momentum, relative to spin angular
momentum, to be in stable equilibrium.

Note that most members of both the \kepler sample considered here and
the set of Hot Jupiters considered previously \citep{levrard} fail to
reside in their tidal equilibrium states. However, the two sets of
planetary systems are dynamically different: The Hot Jupiter systems
generally have too little total angular momentum, so that the
constraint $L>L_X$ is not satisfied; they are thus subject to type-I
instability (see Figure \ref{fig:maplane}).  In contrast, the \kepler
systems considered here generally have enough total angular momentum
(Figure \ref{fig:critang}), but not enough orbital angular momentum
(Figure \ref{fig:hsratio}), and are subject to type-II instability.
This difference arises due to the difference in planetary masses in
the two samples. The \kepler systems generally have much smaller
masses, so that the critical angular momentum $L_X\propto{m}^{3/4}$ is
smaller and the the constraint $L>L_X$ is more easily satisfied. On
the other hand, the orbital angular momentum $h\propto{m}$ is smaller
for these low mass planets, and the constraint $h>3L/4$ is more
difficult to meet.

The considerations of the Section \ref{sec:basic} show that stability 
requires two conditions, which can be written in the form 
\be
{h \over L} > {3 \over 4} \quad {\rm and} \quad L > L_X = 
{4\over3} \left[ {3I G^2 (Mm)^3 \over M+m} \right]^{1/4} \, . 
\label{twocon} 
\ee
To apply these criteria to observed systems, however, we must know
both the orbital angular momentum and the spin angular momentum. The
latter quantity requires additional measurements of stellar properties
(to find the rotation rate) and these are not always available. It is
useful to derive a combined constraint that does not require data for
stellar rotation rates. We can combine the two constraints in equation
(\ref{twocon}) to obtain the weaker condition 
\be 
h > {3 \over 4} L_X = 
\left[ {3I G^2 (Mm)^3 \over M+m} \right]^{1/4} \, . 
\label{combine} 
\ee 
This constraint can be written in the alternate form 
\be
a > \left[ {3 I \over \mumass} \right]^{1/2} \, . 
\ee
We can write the moment of inertia of the star in the form 
$I=\ndi M R^2$, which defines the parameter $\ndi$ (and where 
we expect $\ndi \approx 0.10$ for Solar-type stars). With this
definition, the constraint for stability becomes   
\be
a > \left[ 3 \ndi {M \over \mumass} \right]^{1/2} R \equiv R_C \,,
\label{necessary} 
\ee
where the second equality defines the weighted stellar radius $R_C$.
For example, for Jovian planets, $\mumass/M \approx 10^{-3}$, so the
constraint takes the form $a > R_C \approx 13 R_\ast$. 

\begin{figure} 
\centerline{\psfig{figure=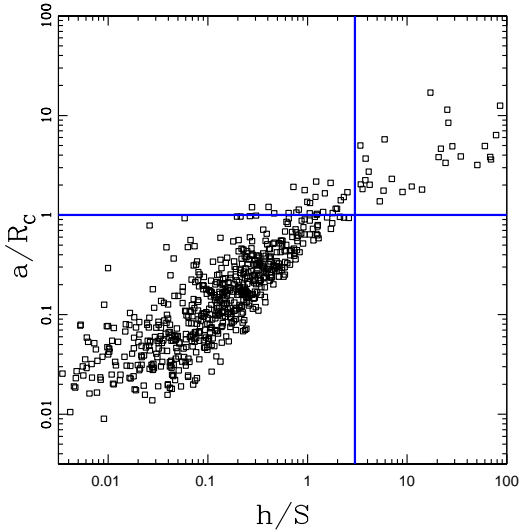,width=8.5cm}}
\caption{Comparison of two stability criteria. The simplified 
condition $a/R_C>1$ is necessary but not sufficient, whereas the 
stricter criterion $h/S>3$ requires that the stellar rotation rates
are measured. Most --- but not all -- of the systems that satisfy the
weaker constraint (and lie above the horizontal line) also satisfy the
stronger constraint (and lie to the right of the vertical line). }
\label{fig:ratrat} 
\end{figure}

The constraint of equation (\ref{necessary}) is necessary but not
sufficient. Any planetary system in equilibrium must satisfy the
constaint, but it remains possible for systems to satisfy the
inequality and still not be in tidal equilibrium. These latter systems
would not satisfy the requirement that $h/S>3$. This issue is
illustrated in Figure \ref{fig:ratrat}, which shows the ratio $a/R_C$
plotted versus the ratio $h/S$ for all of the systems in the sample
that have enough angular momentum to allow for an equilibrium state
(i.e., for the systems with $L>L_X$). The blue lines in the figure
delineate the regions where $h>3S$ and $a>R_C$. The simplified
constraint of equation (\ref{necessary}) does a reasonable job of
specifying the systems that are not in equilibrium. Nonetheless, 24
systems (out of 653) lie above the horizontal blue but do not fall to
the right of the vertical line, i.e., they do not satisfy $h>3S$.

Figures \ref{fig:pstarorb} -- \ref{fig:ratrat} suggest that the
majority of \kepler systems (those in the sample defined above) have
enough angular momentum to allow the existence of a tidal equilibrium
state, but are not actually in a stable equilibrium. Compared to the
conditions required for stability, the orbital periods are not
commensurate with the steller spin periods and the orbital angular
momenta are too small relative to the stellar spin angular momenta.
For most systems, the star spins more slowly than the planet orbits
around it (see Figure \ref{fig:pstarorb}), so that the action of tidal
evolution will move the orbits inward toward the star. Since the
systems already have too little orbital angular momentum, these
planets are scheduled for accretion and hence destruction. For roughly
1/3 of the systems, however, the star spins faster than the orbit, and
tidal torques will act to move the planets outward.

\begin{figure} 
\centerline{\psfig{figure=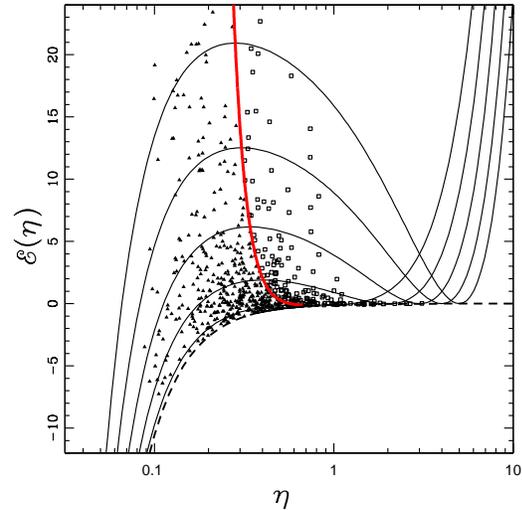,width=8.5cm}}
\caption{Energy versus orbital angular momentum for the sample of  
observed planetary systems. The energy curves are plotted for discrete
values of total dimensionless angular momentum $\ell$ = 1, 2, 3, 4,
and 5 (from bottom to top). The filled triangles depict systems for 
which the star is rotating more slowly than the orbit, wheras the 
open squares depict cases where the star is rotating more quickly. 
The heavy solid red line shows the locus of maxima for the energy 
curves. The heavy dashed curve shows the energy function in the limit 
of zero stellar spin and delineates the lower boundary of the region 
accessible to planetary systems. } 
\label{fig:etadata} 
\end{figure}    

The above point can be illustrated by plotting the observed planetary
systems on the energy curves defined by equation (\ref{reduced}). For
each system, we find the total angular momentum $L$, the critical
angular momentum $L_X$, the ratio $\ell=L/L_X$, and the dimensionless
orbital angular momentum $\eta=h/L_X$. The reduced energy from
equation (\ref{reduced}) is plotted as a function of $\eta$ in Figure
\ref{fig:etadata}. The collection of energy curves shown in the figure
correspond to discrete values of the total system angular momentum
$\ell=L/L_X$ = 1, 2, 3, 4, and 5, where the largest $\ell$ value
produces the largest local maximum. The heavy red solid curve shows
the locations of these maxima for continuous $\ell\ge1$. For systems
in which the star spins faster than the orbital angular velocity, the
points (depicted by open squares) fall to the right of this locus of
maxima.  Similarly, for systems where the star spins more slowly than
the orbit, the points (solid triangles) fall to the left. As a result,
systems that fall to the right of this locus will spiral outward due
to tidal interactions with the star and thereby move toward the tidal
equilibrium state. Systems that fall to the left of the locus must
spiral inward to move toward lower system energies, but no equilibrium
state can be reached. Within the context of this model, the planets
continually spiral inward.  In practice, however, these curves should
be truncated on the left hand side: For a given system, there exists a
minimum orbital angular momentum necessary to keep the planet in orbit
outside the star.

Figure \ref{fig:etadata} shows that more planets will spiral outward
(and hence survive) than the number of planets that already meet the
constraint $h>3I\Omega$ (compare with Figure \ref{fig:hsratio}). This
ordering makes sense: The requirement that the system has enough
orbital angular momentum to evolve toward an equilibrium state is less
stringent than the requirement $h>3I\Omega$ necessary for the system
to reside in a tidal equilibrium state.

We note that the interpretation used throughout this section
implicitly assumes that the stars are spinning in nearly solid-body
rotation with a well defined rate that is measured at the surface. It
remains possible for only the outer layers of the star to participate
in rotation, so that the effective stellar moment of inertia is
smaller than expected, perhaps by a factor of $\sim 10$ (see also
\citealt{levrard}). If this were the case, the spin angular momentum
would be smaller, and more of the systems could be in stable
equilibrium states.

\section{Planetary Systems with Stellar Spin and Quadrupole Moment} 
\label{sec:quad} 

The conditions for tidal equilibrium can be generalized to include the
quadrupole moment of the star. This generalization is interesting for
several reasons. The previous section shows that for one particular
sample, most planetary systems do not reside in a tidal equilibrium
state; we would thus like to know if the inclusion of the quadrupole
moment moves the theoretical equilibrium toward the data (on average).
In addition, the quadrupole term leads to the precession of orbits.
This effect, in turn, affects transit timing variations for orbits
with nonzero eccentricity and/or inclination \citep{agol,holman}, as
well as efforts to measure stellar quadrupole moments \citep{jordi}.
Finally, both the quadrupole and the general relativistic correction
to the potential \citep{hartle} have the same radial dependence
($\Phi\propto1/r^3$), so this derivation informs the corresponding
relativistic problem.\footnote{However, we note that the relativistic  
version of this additional term depends on the angular momentum, so
that the two problems are not equivalent.}

This treatment considers the particular case where the planetary orbit
is circular (so that $e=0$) and lies in the plane defined by the
stellar spin axis (so that $i=0$). Note that these conditions
($e=0=i$) are assumed here, but are required for the tidal equilibrium
states derived in Section \ref{sec:basic}. Notice also that most
planetary candidates with tight orbits tend to have small or vanishing
eccentricity. For systems in this state, the orbital radius and
semimajor axis coincide, so that $r=a$ over the entire orbit. In this
case, the gravitational potential (e.g., see \citealt{hartle}) reduces
to the form 
\be
\Phi = - {G M \over r} - {G \qpot \over r^3} \,.
\label{qpotential} 
\ee
Note that the parameter $\qpot$ can also be written in the form
$\qpot=(1/2)J_2MR^2$, so that $\qpot$ has the same units as the moment of
inertia With this specification, the energy $E$ of the system can 
be written as 
\be
E = - {G M m \over 2a} + {G \qpot m \over 2a^3} + {1 \over 2} 
I \Omega^2 \, . 
\ee
Note that the quadrupole moment of the star will also produce a change
in the stellar moment inertia $I$ and can lead to precession of the
spin axis; for the co-planar systems of interest here, however, we can
ignore precession and simply use the modified value of $I$.  The
angular momentum of the system points in the $\zhat$ direction and has
magnitude $L=h+I\Omega$, where the orbital angular momentum $h$ has
the form 
\be
h^2 = \mumass m \left[ G M a + {3 G \qpot \over a} \right]\,.
\label{horbitq} 
\ee
For circular orbits, this expression follows from the specification of
the potential in equation (\ref{qpotential}) and the definition of
orbital angular momentum (see also \citealt{danby}). 

\subsection{Extremum of the Energy} 

To find the extremum of the energy, subject to the angular momentum
being constant, we find the derivatives of the composite function 
$F = E + \lambda L$ where $\lambda$ is the (single) Lagrange
multiplier in the problem.  We are left with only two variables, 
the semimajor axis $a$ of the orbit and the stellar rotation rate 
$\Omega$.

[1] For the semimajor axis $a$, the derivative 
$\partial F / \partial a$ provides the condition
\be
{GMm \over 2a^2} - {3G\qpot m \over 2a^4} + \lambda 
{\mumass m \over 2h} \left[ GM - {3G\qpot \over a^2} \right] = 0 \,.
\label{dfda} 
\ee

[2] For the stellar spin rate $\Omega$, the derivative 
$\partial F / \partial \Omega$ provides the condition
\be 
I \Omega + \lambda I = 0 \,. 
\ee

The second condition implies that $\lambda = - \Omega$.  Using this
specified value of the Lagrange multiplier, the remaining constraint
of equation (\ref{dfda}) can be written in the form
\be
\left[ {1 \over a^2} - \Omega {\mumass \over h} \right] 
\left[ M - {3\qpot \over a^2} \right] = 0 \,. 
\label{dfda3} 
\ee
Formally, each of the factors could vanish and thereby satisfy the
constraint. However, the parameter $\qpot$ can be written in the form 
$\qpot = (1/2)J_2 M R^2$, where $R$ is the stellar radius and the
dimensionless parameter $J_2 \ll 1$.  We also expect $R\ll{a}$.  
As a result, for almost all realistic systems $M \gg 3\qpot/a^2$, so that
the second factor is nonzero and can be divided out. This leaves us
with the requirement 
\be
h = \mumass \Omega a^2 \,.
\ee 
Using the definition of the orbital angular momentum $h$, 
one finds 
\be
\mumass m \left[ G M a + {3 G \qpot \over a} \right]
= \mumass^2 \Omega^2 a^4 \,,
\ee
which reduces to the form 
\be
\Omega^2 = {G (M + m) \over a^3} 
\left[ 1 + {3 \qpot \over Ma^2} \right]\,. 
\label{syncnew} 
\ee 
In the limit $\qpot\to0$, we recover the previous result from equation
(\ref{sync}).  Even for $\qpot\ne0$, however, equation (\ref{syncnew})
represents synchronous rotation: The correction factor for the stellar
spin rate is the same factor by which the orbital angular velocity
changes with the introduction of a quadrupole moment (for circular
orbits). Finally, we note that the correction to the rotation rate is
${\cal O}(J_2R^2/a^2)\ll1$. In practice, the size of this correction
term will often be much less than the uncertainties in the estimates 
of the stellar masses (which are notoriously difficult to determine). 

\subsection{Minimum Total Angular Momentum} 

The total angular momentum $L=h+I\Omega$. Whereas in the previous case
we could invert the analog of equation (\ref{syncnew}) and write $a$
in terms of $\Omega$, in this case it is easier to eliminate $\Omega$.
After some rearrangement the total angular momentum becomes 
\be
L = \left[ 1 + {3 \qpot \over Ma^2} \right]^{1/2} 
(GMm\mumass)^{1/2} \left\{ a^{1/2} + 
{I \over \mumass a^{3/2}} \right\} \,.  
\ee
We can solve for the critical value of $a$ for which 
the angular momentum is minimized, i.e., 
\be
a^2_{\rm min} = {I \over 2\mumass} 
\left\{ 3 + 3 q + \left[ 9 + 78q + 9q^2 \right]^{1/2} \right\} \,,
\ee
where we have defined the dimensionless parameter 
\be
q \equiv {\mumass \qpot \over MI} \,. 
\label{qdef} 
\ee
Note that $q \ll 1$. Using the critical value of the semimajor 
axis, we can then find the minimum value of the angular momentum,
which can be written in the form 
\be
L_X = \left[ {G^2 (Mm)^3 I \over (M+m)} \right]^{1/4} f(q) \,,
\label{quadlx} 
\ee
where we have defined the dimensionless function 
\be
f(q) \equiv 
\left\{ 5 + 3q + \left[ 9 + 78q + 9q^2 \right]^{1/2} \right\} \qquad
\label{fdefine} 
\ee
$$
\times 
{ \left\{ 3 + 9q + \left[ 9 + 78q + 9q^2 \right]^{1/2} \right\}^{1/2}  \over
2^{1/4} \left\{ 3 + 3q + \left[ 9 + 78q + 9q^2 \right]^{1/2} \right\}^{5/4} }\,.
$$
Since $q \ll 1$, we can find the leading order correction. 
After expanding we thus obtain  
\be
f(q) = {4 \over 3^{3/4}} 
\left\{ 1 + {1\over2} q + {\cal O} (q^2) \right\}\,.
\ee
In the limit $q\to0$, the function $f(q)$ reduces to 
\be
f(q) \to {4 \over 3^{3/4}} \,. 
\ee
As a result, in this limit we recover the solution obtained previously
for the case where the potential has no quadrupole term (as expected). 

\subsection{Second Variation} 

After eliminating the stellar spin variable, we can write 
the energy of the system in the form 
\be
E(a) = - {GMm \over 2a} + {G\qpot m \over 2 a^3} + 
{1 \over 2I} \left( L^2 + h^2 - 2Lh \right) \,,
\ee
where $L$ is the total angular momentum and $h$ is the orbital 
angular momentum given by equation (\ref{horbitq}). The first 
derivative can be written
\be
{dE\over da} = 
{GMm \over 2a^2} - {3G\qpot m \over 2 a^4} + {1 \over I}
\left( h - L \right) {dh \over da} = 0 \,,  
\ee
and the second derivative becomes 
\be
{d^2E\over da^2} = 
- {GMm \over a^3} + {6 G\qpot m \over a^5} 
\ee
$$
+ {1 \over I} \left[ \left( {dh \over da} \right)^2 
+ (h-L) {d^2 h \over da^2} \right] \,. 
$$
After some algebra, we can write the requirement that 
the second derivative is positive in the form 
\be
\left[- \left( 1 + 3q {L-h \over h} \right) + {L \over 4(L-h)} 
\left( 1 - 3q {L-h \over h} \right) \right] 
\ee
$$
\times \left[ 1 - 3q {L-h \over h} \right] > 0 \, ,
$$
where we have used the dimensionless parameter $q$ defined by equation
(\ref{qdef}). The second factor is positive for small $q$, but in
general requires the (rather weak) constraint 
\be
h > {3q \over 1 + 3q} L \, . 
\label{constone} 
\ee
Positivity of the remaining factor requires that 
\be 
hL(1+3q) - 3qL^2 > 4(L-h) \left[ h (1-3q) + 3qL \right]\,,
\ee
which can be solved to find the constraint 
\be
{h \over L} > g(q) \,, 
\label{consttwo} 
\ee
where the function $g(q)$ is defined as 
\be
g(q) \equiv 
{3(1-9q) + \left[9(1-9q)^2+240q(1-3q)\right]^{1/2} \over 8(1-3q)} \,.
\label{gdefine} 
\ee
To leading order, we can write 
\be
{h \over L} > {3 \over 4} 
\left\{ 1 + {2 \over 3} q + {\cal O}(q^2) \right\}\,.
\label{contwolead} 
\ee
In the limit $q\to0$, this constraint reduces to the now-familiar 
form $h/L > 3/4$. Note that the constraint of equation (\ref{consttwo})
is much more stringent than that of equation (\ref{constone}). 

Equations (\ref{consttwo} -- \ref{contwolead}) show that the inclusion
of the quadrupole term results in a tighter constraint: A larger 
fraction of the total angular momentum must reside in the orbit. For
the observational sample of Section \ref{sec:apply}, however, this
corrrection is small.  The parameter $q$ = $(\mumass/M)(\qpot/I)$ =
$(\mumass/M)(J_2/2)$. The first factor is typically
$\mumass/M\sim10^{-4}$ and the second factor $J_2/2 \sim 10^{-6}$, so
the correction is not large enough to move the observed planetary
systems out of (or into) equilibrium states. 

For completeness we note that the results of this section can be
useful in other settings. For example, the dwarf planet Haumea has a
large effective $J_2\approx0.24$ \citep{ragozzine} due its rapid spin
and asymmetrical shape, where this $J_2$ value is obtained by
time-averaging a rapidly rotating ellipsoid with the observed axis
ratios.  In addition, the dwarf planet has two large satellites,
Namaka and Hi’iaka, which have experienced an interesting dynamical
history \citep{cuk}. The results of this section can be used to
describe the tidal equilibrium states of this system, and others.

\subsection{Size of the Planetary Spin Term} 

The treatment thus far has neglected the spin of the planet. In the
equilibrium state, the planet is expected to have a spin rate that is
synchronous with its orbital angular velocity. It is useful to compare
the size of the additional energy term due to the quadrupole moment
with that due to the planetary spin. We can write the ratio in the form 
\be
\ratqs = {G\qpot m \over 2a^3} {2 \over I_P \Omega^2} 
= {\qpot m \over I_P M} \,.
\ee
If the planet and the star have the same internal structure, then 
$I_P = (m R_P^2/MR^2) I$, where $I$ is the stellar moment of inertia. 
The ratio of energy terms then becomes 
\be
\ratqs = {\qpot \over I} \left( {R \over R_P} \right)^2 
\approx 100 {\qpot \over I} \,. 
\ee
It is thus possible for the spin term to dominate the quadrupole term,
or for the two terms to have comparable sizes ($\ratqs\sim1$).

Fortunately, we can readily incorporate planetary spin into our
previous results: Since the extremum condition requires that both
bodies and the orbit have the same angular velocity, the rotation
energy and angular momentum are generalized to the forms 
\be
K_R = {1 \over 2} I \Omega^2 + {1 \over 2} I_P \Omega^2 
\quad {\rm and} \quad 
L_S = I \Omega + I_P \Omega \,, 
\ee
where $I$ is the stellar momentum of inertia and $I_P$ is the
planetary moment of inertia. As a result, we can incorporate the
effects of planetary spin by making the substitution 
\be
I \to I + I_P \,. 
\ee 
This generalized expression should be used for the moment 
of inertia in equations (\ref{deflx}) and (\ref{quadlx}). 
 
\section{Time Evolution} 
\label{sec:evolve} 

Even with the generalization of the stability criteria considered in
Sections \ref{sec:quad}, the majority of the systems in the sample are
not in a tidal equilibrium state. One explanation for this finding is
that the systems are still evolving, but the relevant time scales are
long. More specifically, we expect tidal interactions between the
planets and their host stars to cause them to spiral inward (outward)
when the stellar spin is slower (faster) than the angular velocity of
the orbit. For most systems, the orbital angular speed is larger than
the stellar spin (Figure \ref{fig:pstarorb}) so that planets are
expected to be accreted in the long run.  Because the planets are
still observable, however, the time scale for them to spiral into
their stars must be longer than the typical age of the systems.

To test this hypothesis, we consider the time required for a planet to
spiral inward due to tidal dissipation in the star. Because the
coefficient that sets the magnitude of this effect, the value of 
$Q_\ast$, is highly uncertain, we consider a simple model where the 
semimajor axis of the planet evolves according to 
\be
{d \over dt} \left( {a \over a_0} \right) = - \tau_\ast^{-1} 
\left( {\omegaorb - \Omega \over \omegaorb} \right) 
\left( {a \over a_0} \right)^{-11/2} \,. 
\label{dadt} 
\ee
This form follows from previous work (e.g., \citealt{goldsoter,hut1981}). 
In particular, we start from equation (3) of \cite{levrard} and take
the limiting form where the eccentricity is zero and the system is
aligned.  The factor $(\omegaorb-\Omega)/\omegaorb$ takes into account
the difference between the spin rate of the star and the angular speed
of the orbit (this factor approaches unity when the star spins slowly
compared to the orbit).  The time scale $\tau_\ast$ is given by 
\be
\tau_\ast \equiv 
{2 \over 9} Q_\ast {M \over m} \left( {a_0 \over R} \right)^5 
\left( {a_0^3 \over GM} \right)^{1/2} \,,
\label{timestar} 
\ee
where $a_0$ is the starting value of the semimajor axis and $R$ is the
stellar radius. Note that the definition of the tidal dissipation
parameter $Q_\ast$ varies by dimensionless factors of order unity in
different treatments (e.g., compare 
\citealt{goldsoter,hut1981,adams2006,levrard}). Given the above
definitions, the equation of motion is readily integrated to obtain
the solution 
\be
a(t) = a_0 \left[ 1 - {13\over2} {t \over \tau_\ast} \right]^{2/13} \,,
\label{asolution} 
\ee
where we have taken the limit $(\omegaorb-\Omega)/\omegaorb\to1$. 
The total (possible) evolution time is thus given by $t_{\rm evol}$ =
$(2/13)\tau_\ast$ (see also \citealt{adams2006}).  In practice,
however, the planet will strike the stellar surface at an earlier time
$t=(2/13)\tau_\ast[1-(R/a_0)^{13/2}]$.  The correction is small: For a
typical close planet with a 4-day orbit, the difference between
$t_{\rm evol}$ and the true evolution time is less than one part in a
million. As a result, we can ignore this complication and use $t_{\rm
  evol}$ as the remaining lifetime for the planets. To fix ideas,
consider a planet in a 4-day orbit around a solar type star with
$Q_\ast=10^6$. For a planet with the mass of $m$ = 1 $m_J$ (10
$M_{\earth}$), the evolutionary time $t_{\rm evol}$ $\approx$ 6 Gyr
(180 Gyr). Low mass planets can thus survive a long time in tight
orbits.  This finding suggests the the stellar spin correction (e.g.,
see \citealt{hut1981}) to equation (\ref{dadt}) will generally be
small: For super-Earth-mass planets, in order for the evolution time
to be less than the age of the universe, the orbital period must be
substantially less than 4 days; in contrast, the typical stellar spin
periods are much longer \citep{mcquillan}.

\begin{figure} 
\centerline{\psfig{figure=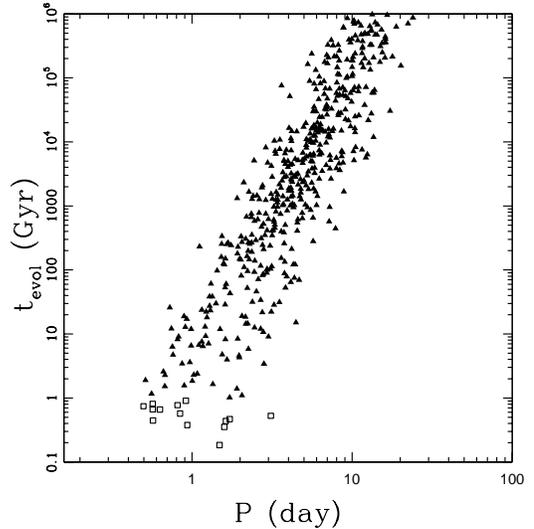,width=8.5cm}}
\caption{Estimated time for planets to spiral into their host 
star (plotted here versus current orbital period). Timescales are 
calculated assuming $Q_\ast=10^6$. } 
\label{fig:time} 
\end{figure}     

Figure \ref{fig:time} shows the time scales $t_{\rm evol}$ for our
observed sample of planets to spiral in to their host stars according
to equations (\ref{dadt}), (\ref{timestar}), and (\ref{asolution}).
These time scales, which are given in Gyr, are plotted versus the
current orbital periods.  This figure shows the resulting time scales
for the subsample of systems that are unstable and have stellar
periods longer than the orbital periods (if this second criterion is
not met, the planets would spiral outwards). For completeness, we
include the correction for the fact that the stellar rotation rate is
not zero \citep{hut1981}, but it has little effect on the results.
For the sake of definiteness, Figure \ref{fig:time} is constructed
using the typical value $Q_\ast = 10^6$ for the tidal quality
parameter of the star. With this value for $Q_\ast$, the majority of
planetary systems have lifetimes in excess of 1 Gyr; only 14 of the
planets (2.6\% of the subsample) are scheduled to be accreted within
this time. For comparison, the ages of the stars are typically in 
the range 1 -- 6 Gyr. This finding is sensible: One expects a small 
fraction of the planets to have short remaining lifetimes. 

\begin{figure} 
\centerline{\psfig{figure=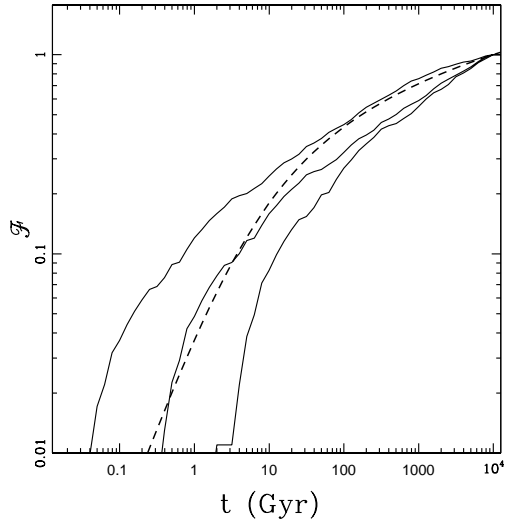,width=8.5cm}}
\caption{Distribution of decay times for the planet sample. 
For the observed planets in the sample, the solid curves show the
cumulative probability distribution for the time required to spiral
into the host star (limited to $t\le10^4$ Gyr). The three curves show
the projected results for different choices of the stellar tidal
dissipation parameter $Q_\ast$ = $10^7$ (top), $10^6$ (middle), and
$10^5$ (bottom). The dashed curve shows the expected cumulative 
distribution for an ensemble of systems with an initial distribution 
that is uniform in $\log(t)$ with ages in the range 1 -- 6 Gyr. } 
\label{fig:tprob} 
\end{figure}     

Although the small number of planets with small projected survival
times seems sensible, we can test this idea further by constructing
probability distributions. In order to proceed, however, we must
specify the starting distribution, which is unknown. The current
observational data-base for extrasolar planets shows that the
distribution of orbital periods or semimajor axis is log-random to
leading order. As a benchmark example, we thus suppose that the
starting distribution of survival times $t_0$ is log-random, i.e.,
\be
{dP \over d\xi} = \norm = constant \qquad {\rm where} \qquad 
\xi \equiv \log t_0 \,,
\ee
where $t_0$ is the time required for a planet to spiral into its host
star from its starting orbit and where $\norm$ is the normalization
constant.  If $\Delta t$ is the age of the system, then the time
required to spiral into the star from the current configuration is
given by 
\be
t = t_0 - \Delta t \,.
\ee
Next we define 
\be
\eta = \log t = \log \left[ t_0 - \Delta t \right]\,,
\ee 
and find the probability distribution for the current set of
survival times 
\be
{dP \over d\eta} = \norm { {\rm e}^\eta \over {\rm e}^\eta + \Delta t} = 
\norm { t \over t + \Delta t} \,.
\ee
This expression would provide the probability distribution if all of
the systems have the same age $\Delta t$. We can include an age spread
by assuming that the system ages are uniform distributed between a
mininum age $\tau_1$ and a maximum age $\tau_2$. The observed age 
distribution for \kepler-planet-hosting stars extends from 
roughly $\tau_1 \approx 1$ Gyr to $\tau_2 \approx 6$ Gyr, although 
it is weighted toward the lower end of the range (\citealt{basri}; 
see also \citealt{lanzashkolnik}). The averaged probability 
distribution then takes the form 
\be
\left\langle {dP \over d\eta} \right\rangle = \norm^\prime 
{t \over \tau_2-\tau_1} \log \left[ {t + \tau_2 \over t + \tau_1} \right] \,,
\ee
where $\norm^\prime$ is the new normalization constant. 
The cumulative probability is then given by 
\be
P (\eta) = \norm^\prime   
\Bigl[ (t+\tau_2)\log(t+\tau_2) 
\label{pdist} 
\ee
$$
- (t+\tau_1)\log(t+\tau_1) - 
\tau_2 \log(\tau_2) + \tau_1 \log(\tau_1) \Bigr] \,. 
$$

Figure \ref{fig:tprob} shows the cumulative distribution of survival
times for both the data and the theoretical considerations outlined
above. Distributions are determined from the observational sample by
using different values for the tidal quality factor $Q_\ast$ = $10^5$,
$10^6$, and $10^7$. As shown in the figure, the distributions obtained
with these values of $Q_\ast$ bracket the probability distributed
calculated using the expression of equation (\ref{pdist}). More
specifically, the data do not favor low values of $Q_\ast\approx10^5$
(cf. \citealt{teitler}), but rather indicate that the tidal quality
factor lies in the range $Q_\ast=10^6-10^7$.

\section{Conclusion} 
\label{sec:conclude} 

This paper has considered the stability of tidal equilibria for
planetary systems with two coupled objectives. First, we have applied
existing stability criteria \citep{hut1980} to a subcollection
\citep{mcquillan} of the candidate planetary systems discovered by the
\kepler mission \citep{batalha}. Second, we have generalized the
classic stability problem to include a quadrupole moment for the
central star. These results indicate that planetary systems are
generally not in --- or near --- tidal equilibrium states.

\subsection{Summary of Results} 
\label{sec:summary} 

We have applied existing stability criteria to observed exoplanet
candidates (Section \ref{sec:apply}) by considering the innermost
planet and the star as the planetary system. Stability requires that
the orbital period is the same as the stellar rotational period
(equation [\ref{sync}]; \citealt{hut1980}). Observed systems are
generally far from synchronous (Figure \ref{fig:pstarorb}); most cases
have longer stellar periods, but some stars are rotating faster than
their planetary orbits.  Stability also requires that the system has
sufficiently large total angular momentum (equation [\ref{deflx}]) and
that at least three-fourths of the angular momentum is contained in
the orbit (equation [\ref{hsthree}]).  We find that most planetary
systems in the sample have enough angular momentum for a stable
equilibrium to exist (Figure \ref{fig:critang}), but that most
planetary systems are not actually in their equilibrium
configuration. The majority of the planetary systems have too little
orbital angular momentum for stability (Figure \ref{fig:hsratio}).

In order for a planetary system to reside in a stable tidal
equilibrium state, it must have enough total angular momentum
(equation [\ref{deflx}]) and enough orbital angular momentum relative
to the total (equation [\ref{hsthree}]). These conditions can be
combined to obtain a necessary --- but not sufficient --- condition
for stability (equations [\ref{combine}] and [\ref{necessary}]). This
new requirement can be evaluated without knowing the spin rate of the
star, and hence can be applied to a wider sample of observed systems.
Moreover, Figure \ref{fig:ratrat} shows that most systems which
satisfy this necessary constraint also meet the more rigorous
constraints.

We have generalized the stability calculation to include the effects
of a stellar quadrupole moment (Section \ref{sec:quad}). Stability
again requires synchronous rotation, where the orbital period includes
the quadrupole correction (equation [\ref{syncnew}]). As before,
stability requires that the total angular momentum is larger than a
threshold value (equation [\ref{quadlx}]) and that a sufficiently
large fraction of the total angular momentum is carried by the orbit
(equation [\ref{consttwo}]). However, this quadrupole correction is
too small to change the conclusion that most of the observed planetary
systems are not in a tidal equilibrium state.

Given that most observed systems (in our sample) are not in tidal
equilibrium, we consider their possible time evolution in Section
\ref{sec:evolve}. These considerations show that the time required for
planets to be accreted onto their host stars is almost always longer
than the age of the system (Figure \ref{fig:time}).  However, some
fraction ($\sim2.6\%$) of the planetary systems have short
evolutionary time scales ($t_{\rm evol}<1$ Gyr) and are expected to be
short-lived. To assess the consistency of this finding, we have
constructed probability distributions for the survival times (Figure
\ref{fig:tprob}). If the tidal quality factor of the stars lies in the
range $Q_\ast=10^6-10^7$, then the observed/inferred distribution of
survival times can be explained with a simple model where the planets
start with a log-random distribution of initial survival times
(basically, a log-random period distribution) and evolve according to
leading order tidal dissipation theory (equations [\ref{dadt}] and
[\ref{timestar}]).

\subsection{Discussion} 
\label{sec:discussion} 

The results of this work apply to the particular sample of 738
candidate planetary systems for which we have measured rotational
periods for the stars \citep{mcquillan}. However, observational biases
could be present. The results presented here assume that the stars are
rotating as solid bodies with their reported rotation rates, and have
a single value of the dimensionless moment of inertia $\ndi=I/(MR^2)$.
The critical angular momentum scales as $L_X \propto I^{1/4}$, so
these variations are not expected to influence the conclusions.
Nonetheless, it remains possible for only the outer layers of the star
to participate in rotation, so that the stellar moment of inertia
would be much smaller. In this case, the star would have less angular
momentum for a given rotation rate, so that the criterion of equation
(\ref{hsthree}) would be more easily satisfied.

The constraints found for the tidal quality factor $Q_\ast$ are also
subject to uncertainties. The evolutionary scenario presented in
Section \ref{sec:evolve} is not unique because the initial period
distribution is not measured and is not necessarily log-random.  The
planetary orbits could also start with nonzero eccentricities, which
would lead to different initial evolutionary paths (although the paths
eventually converge). Additional effects could also influence the
results, including magnetic coupling with the star, stellar braking
through mass-loss, and interactions with additional planets in the
system.  All of these issues should be addressed in future work and as
more data become available. Nonetheless, this paper shows that the
simplest scenario --- a log-random initial period distribution and
$Q_\ast\sim10^6$ --- is consistent with the current \kepler data.

One basic result of this study is that the observed planetary systems
are generally not in tidal equilibrium states. Moreover, most systems
are apparently far from equilibrium. Tidal equilibrium implies
synchronous states, but Figure \ref{fig:pstarorb} shows that stellar
rotation periods and orbital periods are generally not equal and are
not even strongly correlated. Stability also requires that the orbital
angular momentum exceed the spin angular momentum by a factor of 3,
but Figure \ref{fig:hsratio} shows that most systems fall short of
this benchmark by factors of 10 to 100. In the presence of dissipation
--- from any source, not only the simple tidal model encapsulated by
equation (\ref{dadt}) --- one expects systems to evolve toward an
equilibrium state. 

Taken together, the findings summarized above argue for two properties
of these planetary systems. The first property is that any coupling
between the stars and the planets is weak. Typically, these systems
are dynamically old, having experienced of order $10^{11}$ orbits, and
yet they still are not even close to an equilibrium state.  The second
property is that the formation mechanism is largely independent of
tidal equilibrium considerations. The planetary systems are produced
so far from their tidal equilibrium states that they almost always
remain far away after $\sim10^{11}$ orbits.

\medskip 
\textbf{Acknowledgments:} We would like to thank Konstantin Batygin
and Norm Murray for useful discussions. We also thank an anonymous
referee for constructive criticism that improved the paper.  AMB is
supported by grants NSF-DMS-1207693 and INSPIRE-1343720. Part of this
work was carried out during the 2014 International Summer Institute
for Modeling in Astrophysics (ISIMA), which focused on gravitational
dynamics, and was hosted by the Canadian Institute for Theoretical
Astrophyics (CITA).  We are grateful for the hospitality and resources
of CITA, ISIMA, and Univ. Michigan.

\label{lastpage} 

\end{document}